\documentclass[12pt,a4paper]{article}
\usepackage{amsmath,amssymb,amsfonts}
\usepackage{graphicx,a4wide}
\usepackage[top=2cm,bottom=2cm,left=0.5cm,right=0.5cm]{geometry}

\newcommand{\be}{\begin{equation}}
\newcommand{\ee}{\end{equation}}
\newcommand{\bea}{\begin{eqnarray}}
\newcommand{\eea}{\end{eqnarray}}
\newcommand{\bm}{\bibitem}

\newcommand{\al}{\alpha}
\newcommand{\gm}{\gamma}
\newcommand{\Gm}{\Gamma}
\newcommand{\ep}{\epsilon}
\newcommand{\de}{\delta}
\newcommand{\De}{\Delta}
\newcommand{\om}{\omega}

\newcommand{\tht}{\theta}

\newcommand{\lm}{\lambda}

\newcommand{\sg}{\sigma}

\newcommand{\oq}{\overline{q}^{\,2}}

\newcommand{\dmd}{\partial_\mu}
\newcommand{\dnd}{\partial_\nu}
\newcommand{\dmu}{\partial^\mu}

\newcommand{\rt}{\sqrt 2}
\newcommand{\emn}{\epsilon_{\mu\nu\lambda\sigma}}

\newcommand{\op}{\overline\Pi}
\newcommand{\iop}{{\rm Im}\,\overline\Pi}
\newcommand{\og}{\overline G}  
\newcommand{\wt}{\widetilde}
\newcommand{\tom}{\widetilde{\om}}   
\newcommand{\vpamu}{v_\mu + a_\mu}
\newcommand{\vmamu}{v_\mu - a_\mu}

\newcommand{\vk}{\vec k}

\newcommand{\vq}{\vec q}
\newcommand{\vx}{\vec x}
\newcommand{\la}{\langle}
\newcommand{\ra}{\rangle}
\newcommand{\bk}{\boldsymbol{k}}

\newcommand{\rw}{\rightarrow}
\newcommand{\mn}{\mu\nu}
\newcommand{\del}{\partial}

\newcommand{\F}{F_\pi}
\newcommand{\cl}{\cal {L}}
\begin{document}

\setcounter{page}{1}

\title{Analytic structure of $\rho$ meson propagator at finite temperature}

\author{Sabyasachi Ghosh$^{1}$, S. Mallik$^{2}$ and Sourav  Sarkar$^{1}$}
\maketitle
\begin{center}
$^{1}$Variable Energy Cyclotron Centre, 1/AF, Bidhannagar, 
Kolkata, 700064, India\\
$^{2}$Theory Division, Saha Institute of Nuclear Physics, 1/AF 
Bidhannagar, Kolkata 700064, India
\end{center}

\abstract{We analyse the structure of one-loop self-energy graphs for the
$\rho$ meson in real time formulation of finite temperature field theory. We find the 
discontinuities of these graphs across the unitary and the Landau cuts. These contributions 
are identified with different sources of medium modification discussed in the literature. 
We also calculate numerically the imaginary and the real parts of the self-energies and 
construct the spectral function of the $\rho$ meson, which are compared with an earlier 
determination. A significant contribution arises from the unitary cut of the $\pi\om$ loop, 
that was ignored so far in the literature.}

%\pacs{11.10.Wx}

%\keywords{}

\section{Introduction}

The in-medium propagation of vector mesons, particularly the $\rho$, has
been extensively studied in the literature, as reviewed in Refs.\cite{Rapp1,Sarkar}.
The reason is, of course, that it controls the rates of dileptons and photons 
emitted from the hot and dense matter, created during heavy ion collisions. 
The recent precision measurement \cite{Specht} of the in-medium $\rho$
spectral function encourages further theoretical investigation.

The effect of medium on the vacuum propagation of the vector meson is generally 
believed to arise from two sources \cite{Rapp1}. One is the change in its pion
cloud, given essentially by the $\pi\pi$ self-energy loop \cite{Kapusta}. The 
other is the collisions suffered by the vector meson with particles in the
medium. The latter effect can be calculated, broadly speaking, in two different ways. 
Kinetic theory expresses this collision rate in terms of the spin average of the 
squared scattering amplitudes \cite{Haglin}. This rate is simply related to the imaginary 
part of the self-energy of the meson \cite{Weldon}. The effect of collisions can 
also be obtained from the self-energy tensor given by the virial formula, which relates 
it directly to the scattering amplitude itself \cite{Leutwyler1,Eletsky,Rapp2,Jeon,Mallik}. 
In the absence of scattering data, these amplitudes are generally calculated from the 
relevant Feynman graphs. 

As shown by Weldon and others \cite{Weldon,Kobes}, the two sources modifying the free
propagator find a unified description in terms of contributions from branch
cuts of the self-energy loop. 
In addition to the unitary cut, present already in the vacuum amplitude, the thermal 
amplitude generates a new, so-called Landau cut. While the in-medium modification by 
the cloud of virtual particles (mostly pions) is obtained from the unitary cut, 
the effect of collisions with surrounding particles is given by the Landau cut. 
Thus the two sources of medium modification are automatically included in the calculation, 
if we retain the contribution of both the cuts. The relative importance of these cuts 
from different graphs depend on their thresholds, besides the couplings at the vertices 
of the graphs.

In this work we take the one-loop self-energy graphs for the $\rho$ meson
and find all the discontinuities associated with the branch cuts. The loops
are formed with one internal pion line and 
another which may be the pion itself or any of the heavy particles, namely $\omega$, $a_1$ 
and $h_1$, up to a mass of about $1.25$ GeV. The resonances are treated in the narrow width 
approximation. The vertices of the graphs are obtained from chiral perturbation theory. 
The calculations are carried out in the real time version of thermal field
theory.

In Sect. 2 we start with the two-point function at finite temperature of the vector 
current having the quantum numbers of $\rho$. Here we also review briefly the methods 
of chiral perturbation theory to obtain the different interaction Lagrangians needed to 
evaluate the self-energy loops. In Sect. 3 we describe the kinematic decomposition of 
the tensor amplitudes. In Sect. 4 we write explicitly the self-energies from different
loops and separate these analytically into their real and imaginary parts. The cut 
structure of the self-energy function and the discontinuities across the cuts are 
obtained in Sect. 5. In Sect. 6 we evaluate numerically the imaginary and the real 
parts of the self-energy and construct the spectral function of the $\rho$ meson.
Finally Sect. 7 discusses the assumptions and the approximations entering these 
evaluations. Here we also compare our results with an earlier determination. The Appendix  
gives a summary of the real time theory, needed in the present work.

\section{Preliminaries}
\setcounter{equation}{0}
\renewcommand{\theequation}{2.\arabic{equation}}

To study the $\rho$ meson propagator, we do not start directly with the two-point 
function of the $\rho$ meson field, but consider instead that of the vector current, 
having the quantum numbers of the $\rho$ meson. In the two-flavour $QCD$ theory, 
this current is given by 
\be
V_\mu^i(x)=\bar q(x)\gm_\mu\frac{\tau^i}{2}q(x),~~~~~~ q=\left(\begin{array}{c}
u \\ d \end{array}\right),~~~~i=1,2,3 
\ee
Conceptually we then keep contact with the fundamental theory and deal with a 
conserved current in the limit of $SU(2)$ flavour symmetry. At the same time we can 
address directly the physical processes, such as dilepton production in heavy ion 
collisions.

Here we work in the real time formulation of the thermal field theory, where a 
two-point function assumes the form of a $2 \times 2$ matrix. Accordingly we have the 
matrix two-point function
\be
T^{ij,ab}_{\mn}(E,\vq)=i\int d^3xd\tau\,e^{iE\tau-i\vq\cdot\vx}\la
 T_c V^i_\mu(x) V^j_\nu(0)\ra^{ab}
\ee
where $\la{\cal{O}}\ra$ denotes the ensemble average of an operator ${\cal{O}}$,
\be
\la{\cal{O}}\ra=Tr(e^{-\beta H}{\cal O})/Tr e^{-\beta H}
\ee
and  $Tr$ denotes trace over a complete set of states. The superscripts $a,b\,$ 
$(a,b=1,2)$ are thermal indices and $T_c$ denotes time ordering with respect
to a contour in the plane of the complex time variable $\tau$, to be specified in 
the Appendix. However, as we demonstrate there, the basic quantity is again the 
{\it vacuum} two-point function
\be
 T^{ij}_{\mn}(q)=i\int d^4x\,e^{iq\cdot x}\la 0|
 T V^i_\mu(x) V^j_\nu(0)|0\ra \,,
\ee
from which one may easily construct the thermal components. 

In the region of low $q^2$, the $\rho$ meson pole and different low mass branch
points contribute to the two-point function, which may be calculated in
terms of a few hadronic states. As $q^2$ rises, too many hadronic
states start contributing and a continuum sets in. Here the $QCD$ coupling
parameter becomes small, allowing a quark-gluon perturbative calculation of the
two-point function.  

\begin{figure}
\centerline{\includegraphics[width=12cm]{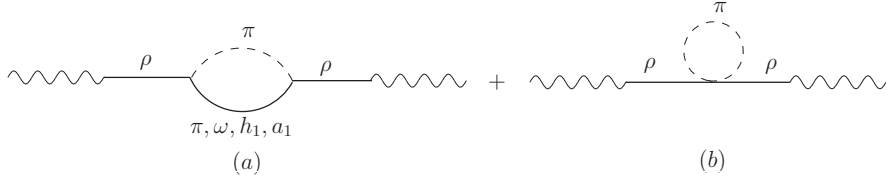}}
\caption{One-loop graphs for the two-point function contributing to
self-energy of $\rho$ meson.}
\end{figure}  

In this work we are interested in finding the leading temperature effect
modifying the free propagation of the virtual $\rho$ meson at $q^2$ below
the continuum. At low temperature, the medium is populated mostly by pions. 
Thus we consider one-loop self-energy graphs of Fig.~1(a), consisting of a pion 
and another low mass hadron $h$, along with the pion itself. There is a
series of such hadronic states with increasing masses. Of course, because of
the presence of thermal distribution function, their contributions fall off
exponentially with their masses at sufficiently high $q^2$. But it is not a
priori clear, up to what mass the hadrons should be retained in our low
$q^2$ calculation. Here we take the particles $h$ to consist of $\om, h_1$ and
$a_1$, postponing discussion on this point to the last Section. Along with
these graphs, we also include the seagull graph of Fig. 1(b), where the loop
represents a single pion propagator, those of heavy particles contributing
insignificantly at low temperatures that we consider here.

The two-point function in the hadronic phase can be evaluated in chiral
perturbation theory by using the method of external fields \cite{Leutwyler2}.
Here one introduces (classical) external vector field $v_\mu^i(x)$ coupled to the vector 
current $V_\mu^i(x)$ (and an axial-vector field $a_\mu^i(x)$ coupled to the
axial-vector current $A_\mu^i(x)$). The resulting terms perturb the original $QCD$ 
Lagrangian as
\be
{\cal L}_{QCD}\to{\cal L}_{QCD}+v_\mu^iV_i^\mu+a_\mu^iA_i^\mu
\ee
The global chiral symmetry group $SU(2)_R\times SU(2)_L$ of massless $QCD$ is thus raised 
to the corresponding local gauge invariance. To arrive at the resulting effective field 
theory, we have to define the various fields and their covariant derivatives
transforming properly under the symmetry group \cite{Leutwyler3,Leutwyler4,Ecker1,Ecker2}. 
The Goldstone fields representing the pions may be described by the unitary matrix,
\be
U(x) = e^{i\pi (x)/\F}~,~~~~~~~~\pi (x)=\sum_{i=1}^3\pi^i(x)\tau^i~~,
\ee
where the constant $\F$ can be identified with the pion decay constant, $\F =93$ MeV and 
$\tau^a$ are the Pauli matrices. One also defines $u(x)$, such that $u^2=U$, because of 
its transformation rule linking $SU(2)_{R,L}$ to the unbroken group $SU(2)_V$. The 
non-Goldstone iso-triplet field $R_\mu (x)$ representing $\rho$ and $a_1$ mesons,
\be
R_\mu (x)=\frac{1}{\sqrt 2}\sum_{i=1}^3\,R_\mu^i (x)\tau^i,
\ee
transforms as
\be
R_\mu\rw h\,R_\mu\,h^\dag,~~~~h(\pi) \in SU(2)_V,
\ee
while the singlet field $S_{\mu}(x)$ representing $\om $ and $h_1$ mesons, of course,
remains unchanged. The covariant derivatives of $U$ and $R_\mu$ are defined by
\be
D_\mu U=\dmd U-i(\vpamu)U+iU(\vmamu),
\ee
and
\be
\nabla_\mu R_\nu=\dmd R_\nu+[\Gamma_\mu,R_\nu]
\ee
with the connection
\be
\Gamma_\mu=\frac{1}{2}\left(u^\dag[\dmd-i(\vpamu)]u+u[\dmd-i(\vmamu)]u^\dag
\frac{}{}\right)
\ee
Finally we have the usual definition of field strengths $F^{\mu\nu}_{R,L}$
associated with the non-abelian external fields $(v_\mu\pm a_\mu)$, where
\(v_\mu=\sum v_\mu^i\tau^i/2\) and \(a_\mu=\sum a_\mu^i\tau^i/2\). We can
now write the leading term in the effective lagrangian for the Goldstone
bosons as
\be
{\cl}=\frac{\F^2}{4}\left\{\la D_\mu U D^\mu U \ra +m_\pi^2 \la
U+U^\dag\ra\right\}
\ee

It is possible to define new variables related to $D_\mu U$ and $F^{\mu\nu}_{R,L}$, 
namely \cite{Ecker1,Ecker2},
\be
u_\mu=iu^\dag D_\mu U u^\dag =u_\mu^\dag
\ee
and
\be
f_\pm^{\mn}=\pm u^\dag F_R^{\mn} u + u F_L^{\mn} u^\dag
\ee
which transform exactly as $R_\mu(x)$ in (2.8). We now have all the variables relevant 
to non-Goldstone bosons, that transform under $SU(2)_V$ only. It is now
simple to write the different pieces of the Lagrangian density, that are 
invariant under $SU(2)_V$ and hence also under $SU(2)_R\times SU(2)_L$. 
Accordingly, we have the kinetic part as  
\be
{\cl}_{kin}=\sum\left( -\frac{1}{4}\la R_{\mn} R^{\mn}\ra + \frac{1}{2}m_R^2
\la R_\mu R^\mu \ra \right) +\sum\left( -\frac{1}{4}S_{\mn} S^{\mn} +
\frac{1}{2} m_S^2 S_\mu S^\mu\right)\,,
\ee
where
\be
R_{\mn}=\nabla_\mu R_\nu-\nabla_\nu R_\mu,~~~~S_{\mu\nu}=\dmd S_\nu -\dnd S_\mu .
\ee
and the two sums run over the iso-triplets and the iso-singlets. Here and below the 
symbol $\la A \ra$ stands for the trace of the 2$\times$2 matrix $A$. The interaction 
vertices to leading order in powers of derivatives (momenta) and external fields are given 
by \cite{Ecker1,Ecker2,MS1} 
\be
{\cl}_{int}=\frac{1}{2\rt m_\rho}( F_\rho\la\rho_{\mn}f_+^{\mn} \ra + 
+iG_\rho\la\rho_{\mn}[u^\mu,u^\nu]\ra )
+\frac{g_1}{\rt} \emn \om^\mu\la \rho^{\nu\lm}u^\sg \ra
+ \frac{g_2}{\rt} h_1^\mu\la \rho_{\mu\nu}u^\nu\ra  
+ \frac{i}{2}g_3\la \rho^{\mn}[a_{1\,\mu},u_\nu]\ra
\ee
Each of these interaction terms is typically a series in powers of the 
derivative of the pion field. Retaining terms to lowest order, we get the
interaction vertices as,
\bea
{\cl}^{(1)}_{int}&=&\frac{F_\rho}{m_\rho}
\del^\mu\vec v^\nu\cdot(\del_\mu\vec\rho_\nu-\del_\nu\vec\rho_\mu)\nonumber\\
&&-\frac{2G_\rho}{m_\rho\F^2}
\del_\mu\vec\rho_\nu\cdot\del^\mu\vec\pi\times\del^\nu\vec\pi\nonumber\\
&&+\frac{g_1}{\F}\ep_{\mu\nu\lm\sg}(\del^\nu\om^\mu\vec\rho^\lm-
\om^\mu\del^\nu\vec\rho^\lm)
\cdot\del^\sg\vec\pi\nonumber\\
&&-\frac{g_2}{\F}h_1^\mu(\del_\mu\vec\rho_\nu-\del_\nu\vec\rho_\mu)\cdot
\del^\nu\vec\pi\nonumber\\
&&+\frac{g_3}{\F}(\del_\mu\vec\rho_\nu-\del_\nu\vec\rho_\mu)\cdot\vec
a_1^\mu\times\del^\nu\vec\pi
\eea
The magnitude of the coupling 
constants may be determined from the observed decay rates of the particles \cite{PDG}. 
Thus the decay rate $\Gm (\rho^0\to e^+\,e^-)= 6.9 $ keV gives $F_\rho =154$ MeV. The 
decay rate  $\Gm (\rho \to 2\pi) = 153$ MeV gives $G_\rho = 69$ MeV. 
Similarly the decay rates $\Gm (\omega \to 3\pi) = 7.6$ MeV,\,  $\Gm (h_1 \to \rho\pi) 
\simeq  360$ MeV and $\Gm (a_1\to \rho\pi) \simeq 400$ MeV give respectively
$g_1=.87,\, g_2=1.0$ and $g_3=1.1$.

The leading term for the four-point vertex needed in the seagull graph (Fig. 1(b))
arises from the kinetic term for the $\rho$ meson appearing in (2.15),
\be
{\cl}^{(2)}_{int}= -\frac{1}{8\F^2}\la
[\rho^\nu, \dmd\rho_\nu][\pi,\dmu\pi]\ra
\ee
We note that this term in the chiral Lagrangian reproduces the famous current algebra 
result for the leading threshold behaviour of pion scattering off a heavy target, which is 
$\rho$ in the present case \cite{Weinberg1}.
 
\section{Kinematics}
\setcounter{equation}{0}
\renewcommand{\theequation}{3.\arabic{equation}}

The isospin structure of $T^{ab,ij}_{\mn}$ is clearly given by $\de^{ij}$,
which we omit from now on. We need to analyse its thermal and Lorentz index structures 
to extract the relevant scalar functions.

We introduce the (four-dimensionally) transverse $\rho$ meson propagator $G_{\mn}^{ab}$ 
by removing from $T_{\mn}^{ab}$ the factor $K \equiv (F_\rho q^2/m_\rho)^2$, representing 
the current-$\rho$ coupling at both ends of graphs of Fig.~1,
\be
T_{\mn}^{ab}(q)=K G_{\mn}^{ab}(q)
\ee
The free elements are 
\be
T_{\mn}^{(0)ab}(q)=K G_{\mn}^{(0)ab}(q),~~~ G_{\mn}^{(0)ab}(q)
=\left(-g_{\mn}+\frac{q_\mu q_\nu}{q^2}\right) D^{ab}(q),
\ee
where $ D^{ab}(q)$ is the free thermal propagator of a scalar particle of mass $m_\rho$.
Comparing with  Eq.(A.15), we find that $G_{\mu\nu}^{(0)ab}(q)$ is indeed the 
(four-dimensionally) transverse part of the free vector propagator $D_{\mu\nu}^{ab}(q)$. 

Now $G$ satisfies the familiar Dyson equation,
\be
G_{\mn}^{ab}(q)=G_{\mn}^{(0)ab}(q)- G_{\mu\lm}^{(0)ac}(q)\Pi^{cd,\lm\sg}(q)
 G_{\sg\nu}^{db}(q)
\label{dyson_G}
\ee
where $\Pi_{\mn}^{ab}(q)$ represents the $\rho$ meson self energy. As shown in the Appendix, 
we can diagonalise the thermal matrices to get rid of the thermal indices, getting the 
Dyson equation in terms of the diagonal elements (denoted by a bar) as
\be
\og_{\mn}(q)=\og_{\mn}^{(0)}(q)-\og_{\mu\lm}^{(0)}(q)\op^{\lm\sg}(q)
\og_{\sg\nu}(q), ~~~~~\og^{(0)}_{\mn}(q)=\left(-g_{\mn}+
\frac{q_\mu q_\nu}{q^2}\right)\frac{-1}{q^2-m_\rho^2+i\ep}
\label{dyson_G_2}
\ee
Here $\op_{\mu\nu}(q)$ is related to the $11$-component by Eq.~(A.18) of the Appendix.

Being free from thermal indices, the Dyson equation may be solved in a way similar 
to that in vacuum. The vector current conservation leads to
\[ q^\mu\og_{\mn}(q)=q^\nu\og_{\mn}(q)=0 \]
At finite temperature there are two independent second rank tensors satisfying this
condition. To construct them we introduce the four-velocity $u_\mu$ of the heat bath
 (not to be confused with the field $u_\mu$ defined in the previous section), 
even though we shall do explicit calculations in its rest frame 
$[u^\mu=(1,0)]$. We thus have one more scalar variable, $u\cdot q$ in addition to 
$q^2$.  Introducing also the transverse variable $\wt u_\mu=u_\mu-u\cdot q\,q_\mu/q^2$, 
we can build two independent tensors
\be
P_{\mn}=-g_{\mn}+\frac{q_\mu q_\nu}{q^2}-\frac{q^2}{\oq}\wt u_\mu \wt
u_\nu, ~~~~~Q_{\mn}=\frac{(q^2)^2}{\oq}\wt u_\mu \wt
u_\nu,~~~~\oq=(u\cdot q)^2-q^2,
\label{defP+
Q}
\ee
satisfying the projection properties
\be
P\cdot P=-P,~~~~~Q\cdot Q=-q^2 Q, ~~~~~P\cdot Q=0
\ee
While both $P$ and $Q$ are four-dimensionally transverse, $P$ is also
3-dimensionally transverse. In the literature one generally finds the factor
$q^2$ instead of $(q^2)^2$ in the definition of $Q_{\mn}$. However, at finite
temperature dynamical singularity can appear at $q^2=0$. The additional factor of $q^2$ 
keeps the kinematic covariant regular at that point.

We now decompose $\og_{\mn}$ as
\be
\og_{\mn}=P_{\mn}\og_t + Q_{\mn}\og_l
\ee
One can further show that the self energy tensor $\op_{\mn}$ is also
transverse~\cite{Weinberg2}, leading to an identical decomposition,
\be
\op_{\mn}=P_{\mn}\op_t + Q_{\mn}\op_l
\ee
Inserting these decompositions in the Dyson equation (\ref{dyson_G_2}) we
get the solution,
\be
\og_t(q)=\frac{-1}{q^2-m_\rho^2-\op_t(q)},~~~~~
\og_l(q)=\frac{1}{q^2}\frac{-1}{q^2-m_\rho^2-q^2\op_l(q)}
\ee
The scalar self-energies may be obtained from the tensor one by taking trace 
over its indices and contracting them with the velocity four vector,
\be
\op_t=-\frac{1}{2}(\op_\mu^\mu +\frac{q^2}{\bar q^2}\op_{00}),~~~~
\op_l=\frac{1}{\bar q^2}\op_{00} , ~~~\op_{00}\equiv u^\mu u^\nu \op_{\mn}
\label{pitpil}
\ee

Finally we note a kinematic relation between the transverse  and the 
longitudinal components of the propagator at $\vq=0$. As $\vq\to0$, the 
kinematic structures $P_{ij}$ and $Q_{ij}$ depend on how the
limit is reached. This unphysical dependence is eliminated, if we take
\be
\og_t(q_0,\vq=0)=q_0^2\,\og_l(q_0,\vq=0)
\ee
Clearly a similar relation must also hold between $\op_{t,l}$, which is
already implied by Eqs.~(3.9).

\section{Evaluation of loop integrals}
\setcounter{equation}{0}
\renewcommand{\theequation}{4.\arabic{equation}}

One-loop graphs were evaluated in the imaginary time formulation by 
Weldon \cite{Weldon} and subsequently in the real time version by Kobes and Semenoff 
\cite{Kobes}, which we follow to evaluate the graphs at hand. We begin with the 
general form of the self-energy loop {\it in vacuum} with internal lines for pion 
and a hadron $h$,
\be
\Pi_{\mn}(q)=i\int\frac{d^4k}{(2\pi)^4}N_{\mn}(q,k)\De_\pi (k)\De_h(q-k)~,
\ee
where $\De_{\pi,h}(q)$ are vacuum propagators of scalar particles of mass
$m_\pi$ and $m_h$ respectively. Tensor structures associated with the two  
vertices and the vector propagator are included in $N_{\mn}$. Introducing the 
following three gauge invariant (transverse) tensors,
\bea
A_{\al\beta}(q)&=&-g_{\al\beta}+{q_\al q_\beta}/{q^2} ,\nonumber\\
B_{\al\beta}(q,k)&=&q^2 k_\al k_\beta-q\cdot k(q_\al k_\beta+k_\al q_\beta)
+(q\cdot k)^2g_{\al\beta} ,\nonumber\\
C_{\al\beta}(q,k)&=&q^4 k_\al k_\beta-q^2(q\cdot k)(q_\al k_\beta+k_\al
q_\beta) +(q\cdot k)^2q_\al q_\beta .
\eea
we get the tensor $N_{\mn}$ for the different graphs of Fig. 1(a) as 
\bea
N_{\mn}^{(\pi)}(q,k)&=&\left(\frac{2G_\rho}{m_\rho \F^2}\right)^2
C_{\mn}\nonumber\\
N_{\mn}^{(\om)}(q,k)&=&-4\left(\frac{g_1}{\F}\right)^2(B_{\mn}+q^2k^2
A_{\mn})\nonumber\\
N_{\mn}^{(h_1)}(q,k)&=&-\left(\frac{g_2}{\F}\right)^2(B_{\mn}-\frac{1}{m_{h_1}^2}
C_{\mn})\nonumber\\
N_{\mn}^{(a_1)}(q,k)&=&-2\left(\frac{g_3}{\F}\right)^2(B_{\mn}-\frac{1}{m_{a_1}^2}
C_{\mn})
\eea

At finite temperature the 11-component of the corresponding self-energy matrix is 
given by 
\bea
\Pi_{\mn}^{11}(q)&=&i\int\frac{d^4k}{(2\pi)^4}N_{\mn}(q,k)D_\pi ^{11}(k)D_h^{11}(q-k)
\nonumber\\
&=& i\int\frac{d^4k}{(2\pi)^4}N_{\mn}(q,k)\De_\pi (k)\De_h(q-k) \nonumber\\
&&-\int\frac{d^4k}{(2\pi)^3}N_{\mn}(q,k)\{\De_h (q-k) n(\om) \de (k^2-m_\pi^2)
+\De_\pi(k) n(\om')\de ((q-k)^2-m_h^2)\}\nonumber\\
&&-i\int\frac{d^4k}{(2\pi)^2}N_{\mn}(q,k)n(\om) n(\om')\de
(k^2-m_\pi^2)\de((q-k)^2-m_h^2)\,,
\eea
on inserting the expression for the thermal propagators $D_{\pi, h}^{11}(q)$
as given by (\ref{Dlam}). Here $n$ is the distribution function at
temperature $T$ and  $\om$ and $\om'$ are the energies of the pion and the
hadron, $\om=\sqrt{m_\pi^2+\vk^2},\,\,\om'=\sqrt{m_h^2+(\vq-\vk)^2}$. 
Here the first term refers to vacuum. The second and the 
third terms are medium dependent, containing the distribution function linearly and 
quadratically respectively. The third term is purely imaginary. 

It is simple to evaluate the $k_0$ integral of each of the three pieces in (4.4) and 
read off the imaginary part. Then we get $\iop (q)$ on using (A.18), associated with 
the time-ordered product. The imaginary part associated with the retarded propagator 
is also of interest, particularly in writing the dispersion relation for $\op (q)$. 
The two imaginary parts are the same, except for a factor of $\ep (q_0)$. Keeping
the retarded propagator in mind, we write (5.1) below for $\iop (q)$. For the real part 
of $\op (q)$, the medium dependent contribution is obtained from the second term in (4.4),
\bea
{\rm {Re}}\,\op_{\mn} (q)=\int\frac{d^3k}{(2\pi)^3}&& \left\{\frac{n(\om)}{2\om}\left(
\frac{N_{\mn}(k_0=\om)}{(q_0-\om)^2-{\om'}^2}
+\frac{N_{\mn}(k_0=-\om)}{(q_0+\om)^2-{\om'}^2}\right)\right.\nonumber\\
&& \left.+\frac{n(\om')}{2\om'}\left(
\frac{N_{\mn}(k_0=q_0-\om')}{(q_0-\om')^2-\om^2}                                  
+\frac{N_{\mn}(k_0=q_0+\om')}{(q_0+\om')^2-\om^2}\right)\right\}
\eea
where the integrals are principal valued.

The seagull graph (Fig. 1(b)) involves a contraction of two pion fields with
a single derivative at the same space-time point that appear in the interaction  
Lagrangian (2.18). As a result, the thermal components of the resulting self-energy 
tensor has the form
\be
\Pi^{ab}_{\mu\nu}(q)|_{\rm {seagull}} \sim g_{\mu\nu}\int \frac{d^4k}{(2\pi)^4}
q\cdot k\,D^{ab}(k)\,=\,0\,,
\ee
if we recall that $D^{ab}$ is a function of $\bk^2$ and $k_0^2$. Thus in chiral 
perturbation theory, the contribution of each graph of Fig. 1(a) is transverse by 
itself, without requiring the seagull graph of Fig. 1(b) to do so, the latter being 
actually zero.

\section{Analytic structure}
\setcounter{equation}{0}
\renewcommand{\theequation}{5.\arabic{equation}}

In writing the expression for the imaginary part of the self-energy, we convert the 
tensors $N_{\mn}$ and $\op_{\mn}$ into scalars by combining their components as in 
(3.10). Also, throughout this Section, we omit for brevity, the subscripts $t$ and 
$l$ of the scalars, as the equations will be valid for both the transverse and the
longitudinal components. We thus get 
\bea
&&\iop (q_0,\vq)=-\pi\int\frac{d^3\vec k}{(2\pi)^3 4\om\om'}\times\nonumber\\
&&[N (k_0=\om)\{(1+n(\om)+n(\om'))\de(q_0-\om-\om')
-(n(\om)-n(\om'))\de(q_0-\om+\om')\}\nonumber\\
&& + N (k_0=-\om)\{(n(\om)-n(\om'))\de(q_0+\om-\om')
-(1+n(\om)+n(\om'))\de(q_0+\om+\om')\}]
\eea
We shall label the 
four terms above by the indices 1, 2, 3 and 4 in the sequence they appear in (5.1).
The functions $N_{t,l}$ for different graphs may be obtained from (4.3). Here it 
suffices to note that these scalar functions depend on scalars built out of the
four-vectors, $q_\mu,\, k_\mu$ and $u_\mu$, in such a way that they remain invariant 
under the {\it simultaneous} change of sign of $q_0$ and $k_0$. We thus get the
symmetry relations 
\bea
\iop_3(-q)&=&-\iop_2(q)\nonumber\\
\iop_4(-q)&=&-\iop_1(q)\,.
\eea
 
\begin{figure}
\centerline{\includegraphics[width=18cm]{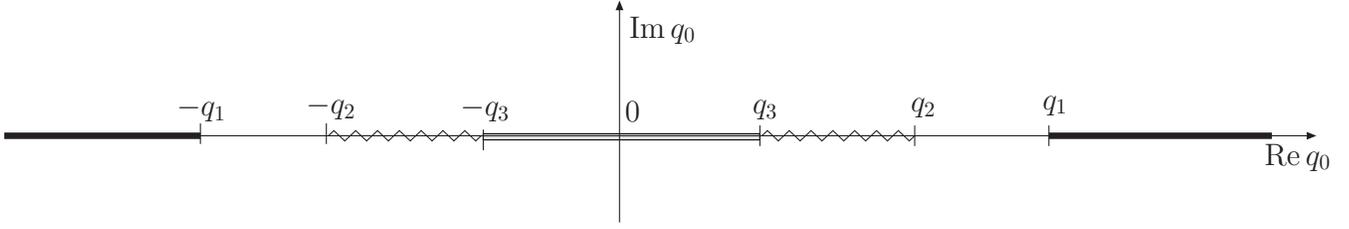}}
\caption{Branch cuts of self-energy function in $q_0$ plane for fixed $\vq$
given by $\pi h$ loop. The quantities $q_{1,2,3}$ denote the end points of cuts 
discussed in the text : $q_1=\sqrt{(m_h+m_\pi)^2+|\vq|^2}$, $q_2=\sqrt{(m_h-m_\pi)^2
+|\vq|^2}$ and $q_3=|\vq|$. For the $\pi\pi$ loop, $q_2$ collapses to $q_3$, 
giving only one variety of Landau cut.}
\end{figure}

The distribution functions present in different terms of (5.1) may be
understood in terms of decay and recombination (inverse decay) probabilities for
processes at the vertices of the loop graphs \cite{Weldon}. Denote $n(\om)$
and $n(\om')$, for short, by $n$ and $n'$. Then  writing, for example,
$1+n+n'=(1+n)(1+n')-nn'$, the first term becomes proportional to the
probability for the decay $\rho \rw \pi +h$ with the expected statistical   
weight $(1+n)(1+n')$ for the stimulated emission {\it minus} the probability
for the inverse decay $\pi +h \rw \rho$ with weight $nn'$ for absorption.
Similarly, writing $-(n-n')=n'(1+n)-n(1+n')$, the second term gives the
probability for $\rho +h \rw \pi$ with weight $n'(1+n)$ {\it minus} that
for $\pi \rw \rho +h$ with weight $n(1+n')$.
 
The regions, in which the four terms of (5.1) are non-vanishing, give
rise to cuts in the self-energy function (Fig. 2). These regions are controlled by the 
respective $\de$-functions \cite{Das}. Thus, the first and the
fourth terms are non-vanishing for $q^2 \geq (m_h +m_\pi)^2$, giving the 
unitary cut, while the second and the third are non-vanishing for 
$q^2 \leq (m_h -m_\pi)^2$, giving the so-called Landau cut. 
The origin of these cuts is clear from the discussion in the last paragraph.
The unitary cut arises from the states, which can communicate with the
$\rho$. These states are, of course, the same as in vacuum, but, as we see above, 
the probabilities of their occurrence in medium are modified by the distribution 
functions. Since it is the distribution function for pions and not for heavy
mesons that dominates the medium dependent probabilities, the 
contribution of the unitary cut is referred in the literature as due to
modification by the pion cloud in medium \cite{Rapp1}. On the other hand, the 
Landau cut appears only in medium and arises from scattering of $\rho$ with 
particles present there. We note that this contribution appears as the first term 
in the virial expansion of the self-energy function \cite{Leutwyler1,Jeon,Mallik}.  
  
We now come back to evaluate (5.1) explicitly, giving discontinuities of
the self-energy function across the cuts in the $q_0$ plane for fixed $|\vq|$. 
We consider a $\pi h$ loop, the $\pi\pi$ loop being a special case.  
Writing $d^3\vk=2\pi\sqrt{\om^2-m_\pi^2}\,\om d\om\,\sin\tht d\tht$, where $\tht$
is the angle between $\vq$ and $\vk$, we can readily integrate over $\cos \theta$ 
using the $\de$-functions. But we have to take into account the physical
requirement, $|\cos \theta| \leq 1$, which, as we shall see presently, reduces
the a priori range ($m_\pi$ to $\infty$) of integration over $\om$. Below we
inspect each of the terms of (5.1) and write down explicitly only those pieces 
that contribute for positive $q_0$.  

Consider the first two terms of (5.1), for which we have
$(q_0-\om)^2=\om^{\prime 2}$, giving
\be
\cos \tht =
\frac{-R^2+2q_0\om}{2|\vq|\sqrt{\om^2-m_\pi^2}}\,,~~~~~R^2=q^2-m_h^2+m_\pi^2\,.
\ee
Then the  inequality $|\cos\tht |\leq 1$ becomes
\be
q^2(\om - \om_+)(\om -\om_-)\leq 0
\ee
where $\om_{\pm}$ are the roots of the quadratic equation for $\om$, 
\be
\om_\pm = \frac{R^2}{2q^2}\{q_0 \pm|\vq|\ep (R^2)v\}\,,~~~
v(q^2)=\sqrt{1 -\frac{4q^2 m_\pi ^2}{R^4}}\,.
\ee

For the first term in (5.1), for which $q^2 \geq (m_h +m_\pi)^2$, as
already stated, we have $R^2 > 0$ and $ v < 1$, so that both $\om_+$ and $\om_-$
have the same sign as that of $q_0$. Then this term is non-zero only 
for positive $q_0$ with the integration variable $\om$ restricted to 
$\om_-\leq \om \leq \om_+$. Changing $\om$ to $x$, $\,
\om=(R^2/2q^2)(q_0+|\vq| x)$,
we get
\be
\iop_1 (q_0,\vq)=-\frac{R^2}{32\pi q^2}\int_{-v}^v dx N(x)
\{1+n(\om)+n(q_0-\om)\}\,,~~~~~~~q_0\geq \sqrt{(m_h+m_\pi)^2+|\vq|^2}\,.
\ee

For the second term in (5.1), we split the region $q^2\leq (m_h-m_\pi)^2$ 
into two segments, namely, $0\leq q^2 \leq (m_h-m_\pi)^2$ and $q^2\leq 0$ with 
positive and negative values of $q^2$, so that the inequality (5.4) may be applied 
immediately. Proceeding as before, we find that the first segment leads to a
cut in the negative $q_0$ region ($-\sqrt{(m_h-m_\pi)^2+ |\vq|^2} \leq q_0
\leq -|\vq|$) with $\om$ restricted to $\om_- \leq \om \leq \om_+$, which
we do not write explicitly. The second segment gives a cut in a region over both 
positive and negative values of $q_0$ with  $\om \geq \om_-$, getting     
\be
\iop_2 (q_0,\vq)=\frac{R^2}{32\pi q^2}\int_v^{\infty} dx N(x)
\{n(\om)-n(\om -q_0)\}\,,~~~~~~~ -|\vq|\leq q_0\leq |\vq|\,. 
\ee

The last two terms of (5.1) may be analysed in the same way. Here the
expressions for $\cos \tht$ and $\om_{\pm}$ remains the same as before, except for
reversal of sign of $q_0$. The position of the cuts in the $q_0$ plane are
thus obtained from the previous ones by reflecting them at the origin. Thus the
third term gives rise to two pieces of imaginary part. With
$\tom=(R^2/2q^2)(-q_0 +|\vq| x)$, we get
\be
\iop_3 (q_0,\vq)=\frac{R^2}{32\pi q^2} \left\{ \begin{array}{ll}
\displaystyle \int_{-v}^v dx N(x)\{n(\tom)-n(q_0+\tom)\}, &  |\vq|\leq q_0\leq
\sqrt{(m_h-m_\pi)^2 +|\vq|^2}\\
-\displaystyle \int_v^\infty dx N(x) \{ n(\tom)-n(q_0+\tom)\}, & -|\vq|\leq q_0\leq |\vq|
\end{array}
\right.
\ee
The fourth term contributes entirely to negative values of $q_0$.

The functions $N_{t,l} (x)$ for different graphs are given by (4.3) with the 
trace and the 00-component of the three tensors calculated from (4.2),
\bea
&& A_\mu^\mu =-3,~~~~ A_{00}=\frac{|\vq|^2}{q^2}\\
&& B_\mu^\mu =m_\pi^2 q^2 +\frac{R^4}{2},~~~~ B_{00}=-\frac{|\vq|^2R^4}{4q^2}(1-x^2)\\
&&  C_\mu^\mu =q^2(m_\pi^2 q^2 -\frac{R^4}{4}),~~~~
C_{00}=\frac{|\vq|^2R^4}{4}x^2
\eea

\section{Numerical evaluation}
\setcounter{equation}{0}
\renewcommand{\theequation}{6.\arabic{equation}}

\begin{figure}
\centerline{\includegraphics[width=8cm]{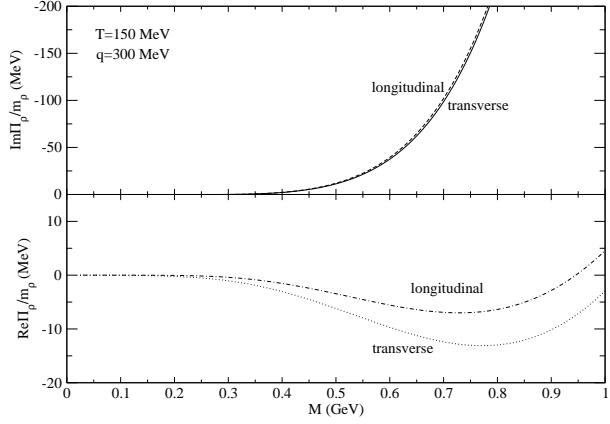}}
\caption{The imaginary and the real parts of self-energy from $\pi\pi$ loop in upper
and lower panel respectively. The longitudinal and transverse components are shown 
separately.}
\end{figure}

\begin{figure}
\centerline{\includegraphics[width=8cm]{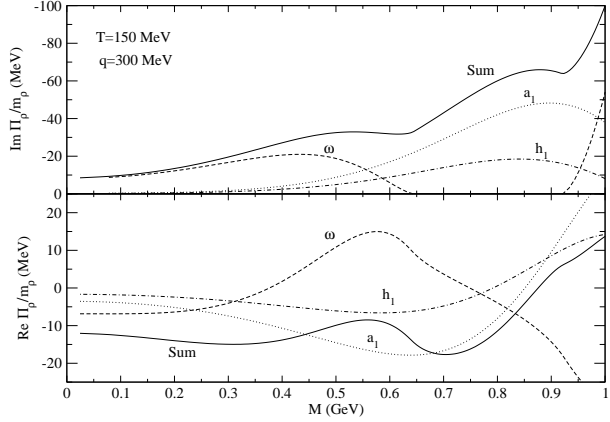}}
\caption{The imaginary and the real parts of self-energy
from different  $\pi h$ loops in the upper and lower panel respectively.
The quantities are averaged over polarisation.}
\end{figure}

\begin{figure}
\centerline{\includegraphics[width=8cm]{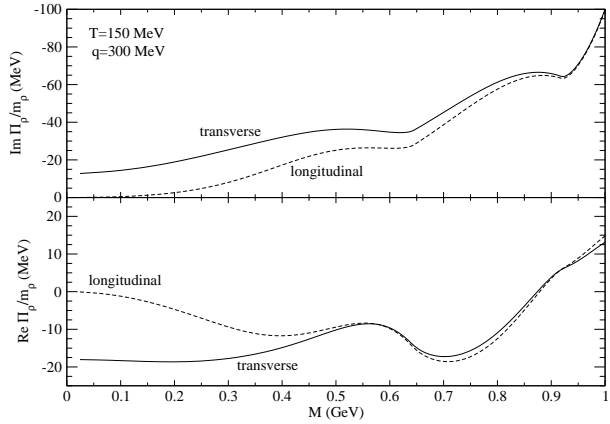}}
\caption{The total imaginary and the real parts of self-energy obtained by
summing over $\pi h$ loops in the upper and the lower panel respectively. The
longitudinal and the transverse components are shown separately.} 
\end{figure}

\begin{figure}
\centerline{\includegraphics[width=8cm]{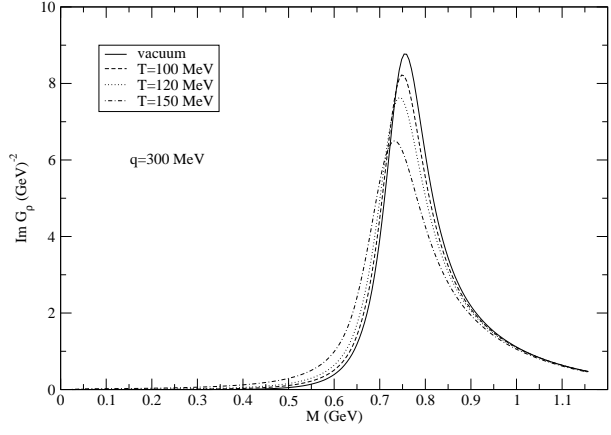}}
\caption{$\rho$ spectral function 
for different temperatures at fixed three-momentum.}
\end{figure}

\begin{figure}
\centerline{\includegraphics[width=8cm]{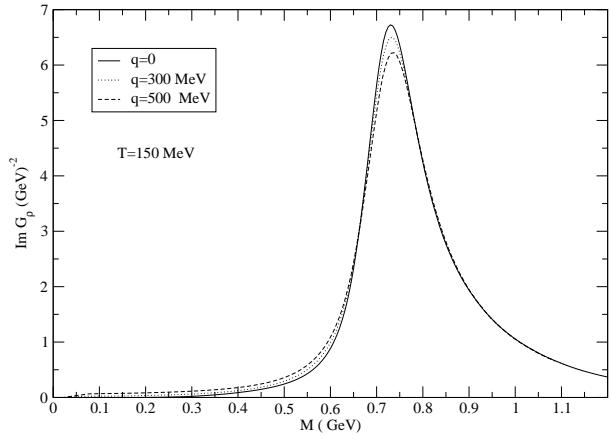}}
\caption{$\rho$ spectral function for different
three-momenta at fixed temperature.}
\end{figure}

We begin with the results of numerical evaluation of the different graphs of 
Fig.~1(a) for the self-energy of the $\rho$. As usual, we retain the vacuum
contribution in the imaginary parts only, assuming the real (divergent) parts
to renormalise the $\rho$ meson mass. We calculate the self-energies as a
function of $\sqrt{q^2}\equiv M$ at fixed values of the three-momentum $\vq$
and temperature $T$. It thus suffices to calculate the self-energies in the 
time-like region, for positive values of $q_0$ starting from $q_0=|\vq|$.
Then the first part of the Landau cut $(0\leq q_0\leq |\vq|)$ cannot appear
in our calculation of the imaginary parts.

The $\pi\pi$ loop is distinguished by a large imaginary part of the self-energy, 
its vacuum part giving $\Gm_\rho\equiv\iop^{(\pi)}/m_\rho=153$ MeV at $M=m_\rho$. 
Clearly it is only the unitary cut in the time-like region that gives the imaginary 
part. The results for this loop are shown in Fig.~ 3.

In showing the results for other loops, we average their imaginary and real
parts over the transverse and longitudinal components,
\be
\op (q)=\frac{1}{3}(q^2\op_l(q)+2\op_t(q))
\ee
They are shown in Fig.~4. Here it is only the second part of the Landau cut
$(|\vq|\leq q_0 \leq \sqrt{(m_h-m_\pi)^2+|\vq|^2})$, which contributes to
the imaginary part. The only exception is the $\pi\om$ loop, where the
unitary cut $(q_0\geq \sqrt{(m_h+m_\pi)^2+|\vq|^2})$ also contributes, its  
threshold for other loops appearing outside the range of $M$ plotted here. 
The $\pi\om$ loop dominates up to about $M\sim 500$ MeV, beyond which the 
$\pi a_1$ loop takes over. The rising trend of the imaginary part at the upper 
end is due to the contribution of the unitary cut. While the imaginary parts add 
up, there is appreciable cancellation among the real parts of different loops.
Fig.~5 shows separately the behaviour of the transverse and the longitudinal 
components of both the real and the imaginary parts of self-energy, summed over 
$\pi h$ loops. Though they differ at low $M$, they tend to converge at higher
$M$.

Finally we come to the transverse and the longitudinal components of the
spectral function, defined by
\be
\mathrm{Im}\,\og_{t,l}(q)=\frac{-\sum\iop_{t,l}}{(M^2-m_\rho^2-
(1,q^2)\sum\mathrm{Re}\,\overline\Pi_{t,l})^2+\{(1,q^2)\sum\iop_{t,l}\}^2}
\ee
where the summation extends over all the loops. Again we take the average
over the two components,
\be
\mathrm{Im}\,\og(q)=\frac{1}{3}(q^2\mathrm{Im}\,\og_l+2\mathrm{Im}\,\og_t)
\ee
It is drawn in Fig.~6 for three different temperatures at $\vq=300$ MeV, 
showing a reduction in the height of the peak with the rise of temperature. 
The pole position, defined by the zero of the real part of of the inverse 
propagator, shifts from $M=775$ MeV in vacuum to $M=763$ MeV at $T=150$ MeV.
Fig.~7 shows this average spectral function for different values of
three-momentum $|\vq|$ at $T=150$ MeV. It appears to depend little on the 
magnitude of $\vq$.

\section{Discussion}

We present a detailed analysis of the singularities of a one-loop, meson
self-energy graph at finite temperature. It is carried out in the real time
formulation of field theory in medium. The branch cuts are obtained in the
(complex) energy plane at fixed three-momentum. The discontinuities across 
the cuts are obtained explicitly. The unitary and the Landau cuts arise from 
what are known in the literature respectively as modifications by pion cloud and in-medium 
scattering, both effects depending largely on the pion distribution function. 
The cut structure unifies the discussion of the two parts of the
self-energy function.   

We also evaluate numerically the self-energy and the resulting spectral
function of $\rho$, which may be compared with an earlier determination by
Rapp and Gale \cite{Rapp2}. Before we do so, however, it would be useful to 
outline the differences in the two treatments of the problem. Of the series 
of resonances with increasing masses, which can contribute in the loop
calculation, we keep only  $\om,\, h_1$ and $a_1$, while they include three
more. Here we note that although we focus on the $\rho$ meson pole with self-energy
corrections, the physical quantity is the complete two-point function, 
consisting of this pole and the continuum. As we take fewer resonances to
calculate the self-energy, we expect the corresponding continuum to begin at
a lower value of $q^2$ than theirs. The dilepton spectra emitted in the
heavy ion collisions depend on the complete two-point function. It would 
thus be of much interest to work out this emission spectrum with these 
different evaluations of the two-point function and compare with experiment.  

They \cite{Rapp2} also include vertex form factors and finite widths of resonances in
their evaluation. The form factors go to decrease the contributions from
higher momenta. So do the distribution functions for calculations in medium.
Thus while the absence of form factors may lead to a significant difference
for results in vacuum, it is not so for their medium dependent parts. The
finite widths referred to are those of resonances inside a loop integral,
where an integration over phase space is involved. In such cases, the
results with finite and narrow widths generally differ by no more
than, say $10 \%$. The case of $\De (1237)$ is an example \cite{Leutwyler1}.
Also approximating the vector and the axial-vector spectral functions
themselves by sharp peaks for $\rho$ and $a_1$ in the well-known Weinberg 
sum rules \cite{Weinberg3,Weinberg4}, one commits an error of only a few 
percent \cite{Donoghue}.  

Finally there is some difference at the level of basic formulae.
Their expression for the real part of the self-energy due to      
singularities in the $q_0\leq0$ region differs from ours (4.5)), even   
though the difference is numerically small. Also the structure of some of   
the interaction Lagrangians are different, but again it would not affect the
results significantly in the neighbourhood of $M=m_\rho$.

Keeping all these differences in mind, we may expect the results of our
calculation to agree with that of ref. \cite{Rapp2} to within, say $20\%$.
Indeed, the imaginary parts agree well within this expectation in a wide
region in $M$ around the $\rho$ meson mass. At higher values of $M$,
however, our calculation shows a substantial increase due to the inclusion of 
the unitary cut of the $\pi\om$ loop. 

To conclude, we derive the discontinuities across all the cuts of a
self-energy loop of a vector meson. In the literature \cite{Rapp2} the
self-energy is constructed from the unitary cut of the $\pi\pi$ loop and the
Landau cut of different $\pi h$ loops. By contrast, we find that the unitary
cut of at least some of the $\pi h$ loops (here $\pi\om$) may also
contribute significantly.

\section*{Acknowledgement}
One of us (S.M.) acknowledges support of Department of Science and
Technology, Government of India.

\section*{Appendix : Real time field theory at finite temperature}
\setcounter{equation}{0}
\renewcommand{\theequation}{A.\arabic{equation}}

In this Appendix we review briefly the real time formulation of thermal field
theory. We construct scalar and vector propagators and diagonalise them,
showing that such matrices are actually given by a {\em single} 
analytic function, coinciding essentially with the corresponding result from
the imaginary time formulation. But the real time procedure requires neither 
the frequency sum to evaluate the loop integrals, nor any analytic continuation 
to physical energies.

\vskip0.2cm
\noindent{\bf Free scalar propagator}
\vskip0.2cm

First consider the (free) propagator for the scalar field $\phi(x)$, 
\be
D(\tau,\tau';\vx-\vx')=i\la T_c\phi(\tau,\vx)\phi(\tau',\vx')\ra\,.
\label{defD_1}   
\ee
where $x^\mu=(\tau,\vx)$. Here $\tau$ and $\tau'$ are any two points on a contour
in the plane of the complex time variable. The contour begins at $-T$ say on the 
real axis and ends at $-T-i\beta$, nowhere moving upwards and remaining entirely
within the analyticity domain $-\beta\le{\rm Im}(\tau-\tau')\le 0$ \cite{Mills}.
Apart from these restrictions we keep the form of the contour arbitrary at this
stage. The delta- and theta- functions on the contour will be denoted by the
subscript $c$.

The thermal propagator satisfies the same differential equation as the one in
vacuum,
\be
(\Box_c+m^2)D(\tau,\tau';\vx-\vx')=\delta_c(\tau-\tau')\delta^3(\vx-\vx'),
\label{eqofmotn}
\ee
but obeys a different, thermal boundary condition. If we write
\be
D(\tau,\tau';\vx-\vx')=\theta_c(\tau-\tau')D^+(\tau,\tau';\vx-\vx')+
\theta_c(\tau'-\tau)D^-(\tau,\tau';\vx-\vx'),
\label{defD_2}       
\ee
and use translation in $\tau$ and cyclicity of the trace in either of
$D^{\pm}$, we get the so-called KMS boundary condition as,
\be
D^-(\tau,\tau';\vx-\vx')=D^+(\tau - i\beta,\tau';\vx-\vx')
\label{kms}
\ee
To obtain the solution it is convenient to take the spatial Fourier transform,
\be
D(\tau,\tau';\vec x-\vec x') = \int\frac{d^3k}{(2\pi)^3}e^{i\vec k\cdot(\vec
x-\vec x')}D(\tau,\tau';\vec k)
\label{D_FT}
\ee
The $\tau$-dependence is then given by 
\be
\left(\frac{d^2}{d\tau^2}+\om^2\right)D(\tau,\tau';\vec k)=\de_c(\tau-\tau')~~~~
~~~\om=\sqrt{\vk^2+m^2}
\label{diffeq_D}
\ee
It is now an elementary exercise in Green's function to solve Eq.(A.6) with
the boundary condition (A.4),
\be
D(\tau-\tau',\vec k)=\frac{i}{2\om}\left[e^{-i\om(\tau-\tau')}
\{\tht_c(\tau-\tau')+n(\om)\}+e^{i\om(\tau-\tau')}
\{\tht_c(\tau'-\tau)+n(\om)\}\right],
\label{solnD_1}
\ee
where $n(\om)=\dfrac{1}{e^{\beta \om}-1}$, getting  
the particle number distribution function in the propagator.

\begin{figure}
\centerline{\includegraphics{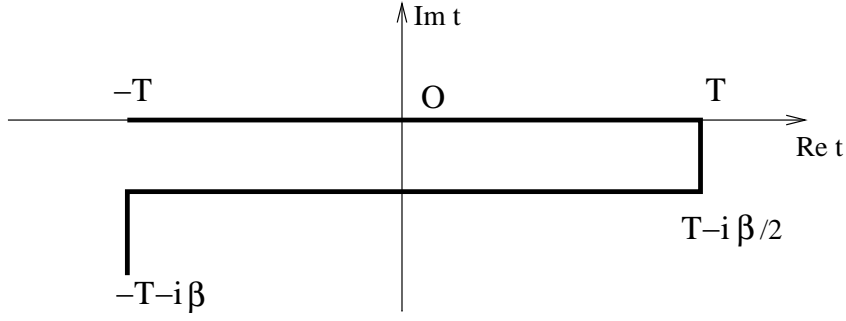}}
%\centerline{\psfig{figure=real_time.eps,height=2.5cm,width=8cm}}
\caption{Contour in time plane for real time formalism}.
\end{figure}

Of the variety of possible contours in the complex time plane \cite{Matsumoto}, 
two are specially interesting, namely the closed one \cite{Keldysh} 
and the symmetrical one \cite{Niemi}. We now choose the latter contour (Fig.~8)
with $T\to\infty$, when $D(\tau-\tau',\vk)$ reduces effectively to four components, 
which may be assembled in the form of a $2\times 2$ matrix. The Fourier transform 
$D^{ab}(k)$  of these components can be defined  with respect to real time to 
give \cite{Niemi,Aurenche,Gelis,MS2}, 
\bea
&&D^{11}=-(D^{22})^*=\De(k)+2\pi in\de(k^2-m^2)\nonumber\\
&&D^{12}=D^{21}=2\pi i\sqrt{n(1+n)}\de(k^2-m^2)
\label{Dlam}
\eea
where $\De(k)$ is the Feynman propagator in vacuum,
\be
\De (k^2)=\frac{-1}{k^2 -m^2 +i\ep}
\ee

The thermal propagator may be diagonalised in the form 
\be
D^{ab}(k_0,\vec k)=U^{ac}(k_0)[{\rm diag}\{\De(k_0,\vk),-\De^*(k_0,\vk)\}]^{cd}
U^{db}(k_0)
\label{diag_D}
\ee
with the elements of the diagonalising matrix as
\[U^{11}=U^{22}=\sqrt{1+n},~~~ U^{12}=U^{21}=\sqrt{n}\]
Writing spectral representations, one can show that $U$ diagonalises not
only the free propagator, but also the complete one \cite{Kobes,MS1}.

\vskip0.2cm
\noindent{\bf Free vector propagator}
\vskip0.2cm

The thermal propagator for a massive, spin-one particle may be derived in a
similar way. Denoting its field by $\rho_\mu(x)$, it has the propagator
\be
D_{\mn}(\tau,\tau';\vx-\vx')=i\la T_c\rho_\mu(\tau,\vx)\rho_\nu(\tau',\vx')\ra\,.
\label{defDmn_1}   
\ee 
satisfying the differential equation
\be
[-g_{\mn}(\Box_c+m^2)+\del_\mu\del_\nu]D^{\nu\rho}(\tau,\tau';\vx-\vx')=
g_\mu^\rho\delta_c(\tau-\tau')\delta^3(\vx-\vx')
\ee
The thermal boundary condition is again given by an equation similar to that
of Eq.(A.4). As before we take the spatial Fourier transform, when the the tensor 
components satisfy
\bea
&&\om^2D^{00}(\tau-\tau',\vk)+ik_l\frac{d}{d\tau}D^{l0}(\tau-\tau',\vk)=
-\delta(\tau-\tau')\nonumber\\
&&\om^2D^{0i}(\tau-\tau',\vk)+ik_l\frac{d}{d\tau}D^{li}(\tau-\tau',\vk)=
0\nonumber\\
&&-\left[g_{il}\left(\frac{d^2}{d\tau^2}+\om^2\right)+k_ik_l\right]D^{lj}(\tau-\tau',\vk)-
ik_i\frac{d}{d\tau}D^{0j}=g_i^j\de(\tau-\tau')
\eea
It is easy to check that these equations as well as the boundary condition
is solved by
\bea
D_{ij}(\tau-\tau',\vk)&=&-\left(g_{ij}-\frac{k_ik_j}{m^2}\right)D(\tau-\tau',\vk)\nonumber\\
D_{0i}(\tau-\tau',\vk)&=&i\frac{k_0}{m^2}\frac{d}{d\tau}D(\tau-\tau',\vk)\nonumber\\
D_{00}(\tau-\tau',\vk)&=&-\left(1+\frac{1}{m^2}\frac{d^2}{d\tau^2}\right)
D(\tau-\tau',\vk)
\eea
where $D$ satisfies the same Eq.~(\ref{diffeq_D}) as for the scalar field.
We now take the temporal Fourier transform as before and assemble the resulting
four dimensional Fourier Transform of the four components of the thermal vector
propagator in terms of that for the scalar propagator
\be
D_{\mn}^{ab}(k)=\left(-g_{\mn}+\frac{k_\mu k_\nu}{m^2}\right)D^{ab}(k)
\ee
the Lorentz structure remaining the same as for the vacuum propagator.

\vskip0.2cm
\noindent {\bf Diagonalisation of complete vector propagator}
\vskip0.2cm

We now turn to Dyson equation (3.3) in the text. As already stated above,
both the free and the complete propagator can be diagonalised by the same
thermal matrix $U$ ; thus
\be
G^{ab}_{\mn}(k_0,\vec k)=U^{ac}(k_0)[{\rm
diag}\{\og_{\mn}(k_0,\vk),\, -\og^*_{\mn}(k_0,\vk)\}]^{cd}U^{db}(k_0)   
\ee  
As a consequence, the matrix $\Pi^{ab}_{\mn}$ is also diagonalisable by
$(U^{-1})^{ab}$, 
\be
\Pi^{ab}_{\mn}(q)=[U^{-1}(q_0)]^{ac}[{\rm
diag}\{\op_{\mn}(q),-\op_{\mn}^*(q)\}]^{cd}[U^{-1}(q_0)]^{db}
\ee
We thus get the Dyson equation (3.4) in the text for the barred quantities,
that are free from the thermal indices.
The diagonalised tensor $\op_{\mn}$ can be related to any of the components of the 
corresponding matrix. For example, it is related to the $11$ component as
\bea
{\rm Re}\,\op_{\mn}&=&{\rm Re}\,\Pi_{\mn}^{11}\nonumber\\
\iop_{\mn}&=&\tanh(\beta \om/2){\rm Im}\,\Pi_{\mn}^{11}
\eea

\end{document}